**Computational and informatics advances for reproducible data analysis in neuroimaging**

Russell A. Poldrack[1*], Krzysztof J. Gorgolewski[2], & Gael Varoquaux[3]

1. Department of Psychology, Stanford University, Stanford, CA USA; russpold@stanford.edu
2. Department of Psychology, Stanford University, Stanford, CA USA; chrisgor@stanford.edu
3. Inria Parietal, Neurospin – CEA, Gif/Yvette, France; gael.varoquaux@inria.fr

* Corresponding Author contact information: Russell Poldrack, Jordan Hall, Stanford, CA 94305-2130. russpold@stanford.edu

# Article Table of Contents

## Abstract:

The reproducibility of scientific research has become a point of critical concern.  We argue that openness and transparency are critical for reproducibility, and we outline an ecosystem for open and transparent science that has emerged within the human neuroimaging community.  We discuss the range of open data sharing resources that have been developed for neuroimaging data, and the role of data standards (particularly the Brain Imaging Data Structure) in enabling the automated sharing, processing, and reuse of large neuroimaging datasets.  We outline how the open-source Python language has provided the basis for a data science platform that enables reproducible data analysis and visualization.  We also discuss how new advances in software engineering, such as containerization, provide the basis for greater reproducibility in data analysis.  The emergence of this new ecosystem provides an example for many areas of science that are currently struggling with reproducibility.

## Introduction

The last decade has seen an avalanche of concern regarding the reproducibility of scientific research, across fields as diverse as psychology (1, 2), cancer biology (3), and economics (4, 5), among others.  While this crisis has inspired a significant degree of hand-wringing, it also has spurred the development of new approaches to improve the reproducibility and transparency of scientific research. In this article we will describe a particular approach to open sharing and reproducible data analysis that has emerged within the brain imaging community, driven in part by this crisis. We focus primarily on the analysis and sharing of data obtained from magnetic resonance imaging (MRI), which is currently the principal method for human neuroscience research (6).  Our goal is to highlight how open practices regarding software and data have transformed the scientific landscape, allowing major advances in knowledge and providing the means to improve scientific reproducibility.

Reproducibility means many different things to different people (7–9).  In the present paper we will focus on the computational aspects of reproducibility, by which we mean that a scientific workflow should be able to produce quantitatively near-identical results when applied to the same input data by the same or other researchers.  We hasten to note that reproducibility is orthogonal to the mathematical or conceptual validity of the results; one can generate highly reproducible but invalid results from a data analysis workflow by simply setting every result to a constant value.  For this reason, we argue that reproducibility must be contextualized by validity, such that a workflow should provide answers that are both reliable and valid in order to be considered to be "reproducible."

The recent increase in the volume of data and computing power used to analyse them implies that the pipelines that produce modern scientific results are increasingly complex, with an increasing number of analytic degrees of freedom. In this context, reproducibility is a growing challenge: Not only is computational reproducibility harder to achieve in the face of greater complexity, but also it matters more, as an increasingly large set of different conclusions may arise from the same data.

**The importance of open science**

In our view openness is central to reproducibility, which implies that other researchers must be able to run the workflow on the same data; for this reason, the use of closed software and/or data that are not available to other researchers leads to irreproducible research by definition. The term "open" is also subject to a broad range of interpretations, so it is useful to specify exactly what we mean by the term and how it is applied. Foremost, we view openness in terms of licensing for reuse, which implies that any research objects should have an explicit license or usage agreement attached. In the context of software, this means the specification of a particular open source license, with a strong preference for minimally restrictive licensing (e.g. using the permissive BSD, MIT, or Apache licenses rather than licenses with restrictions on commercial use or "viral" licenses such as the GNU Public License). We also believe that open platform availability is also essential; for example, while a researcher might release MATLAB code under a open source license, the platform on which the code runs is not open (unless it has been tested to run on the open-source Octave clone), thus limiting the ability of other researchers to reproduce the work. In the context of data, openness means that the data should be shared under a data usage agreement that has the least possible restrictions on reuse while still protecting the confidentiality of the research participants.

In what follows, we outline a set of technical, scientific, and social developments that have contributed to the emergence of an ecosystem for open and reproducible research in neuroimaging. We first outline the emergence of radically open data sharing within the neuroimaging community, and show how the availability of these open data have transformed the ability to study the human brain, particularly through the development of community standards for shared data and "Science-as-a-service" models for data analysis and sharing. We then outline the ways in which open source software developers (particularly via the Python language) have generated a computational infrastructure for transparent research practices, mostly notably in the context of machine learning. We conclude by discussing the importance of software engineering best practices for reproducible data analysis.

# Data sharing in neuroimaging

Data sharing practices vary remarkably across research fields, due to a combination of social and technical factors. The debate over "research parasites" (10–12) provided a stark demonstration of how researchers in some subfields of biomedical science remain stubbornly opposed to wide-scale open data sharing. However, in other subfields including human neuroimaging, the broad and open sharing of data has become relatively common, and in doing so has obviated many of the concerns raised by opponents of data sharing.

Data sharing in the neuroimaging community began with a project known as the fMRI Data Center (fMRIDC), which was founded in 1999 at Dartmouth by Michael Gazzaniga and Jack Van Horn (13).  This project was far ahead of its time, and did not have sufficient community buy-in to become highly successful, though it did share more than 100 datasets until its demise in 2012.  The current era of large-scale open sharing of neuroimaging data was heralded by the 1000 Functional Connectomes Project (organized by the International Neuroimaging Data-sharing Initiative (INDI)), which assembled a large international consortium of 35 imaging groups to share resting state functional MRI (rsfMRI) data from more than 1,400 individuals.  This project allowed the group to amass a dataset that was an order of magnitude larger than any previous dataset, and used these data to demonstrate new features of variability in the human connectome across individuals (14).  Perhaps more importantly, it provided a signal example of open data sharing, which other projects have subsequently followed.  An notable difference between data sharing in neuroimaging and in other fields (such as genomics) is that it has arisen from the ground up, driven by the needs of researchers for larger datasets rather than through mandates or requirements from funding bodies or journals.

The sharing of neuroimaging data has not only increased transparency, but has also resulted in substantial cost savings (15),  For example, de novo data generation required to replicate the publicly shared INDI data would cost roughly $1,000,000,000 (16).  The quality of research performed using publicly available data is comparable to other research, as indicated by the citation counts of the resulting manuscripts. For example in case of the INDI datasets, each study reusing the data accrued on average 20.4 citations.  In addition to new scientific discoveries, shared data have also enabled other uses, such as the development of hands-on courses (17) and the benchmarking of new data analysis methods (18).

## *Varieties of data sharing*

The data that a researcher might share come in various forms, in particular with regard to their "rawness" (i.e. the degree to which they have been processed after acquisition) (see Figure 1). The most robust reproducibility comes from the availability of raw data, because it allows end-to-end reproduction of the data processing and analysis workflow as well as the ability to modify the workflow in order to assess the degree to which results are robust to particular analysis decisions (e.g. "vibration" analysis;(19)).  However, the sharing of raw data also exacts substantial costs due to the size and complexity of the data, and also raises greater concerns regarding the confidentiality of research participants.  In addition, one often wants access to intermediate results, such as images containing the statistical values computed in a workflow, so that a researcher could examine and reuse these without necessarily reproducing the entire workflow.

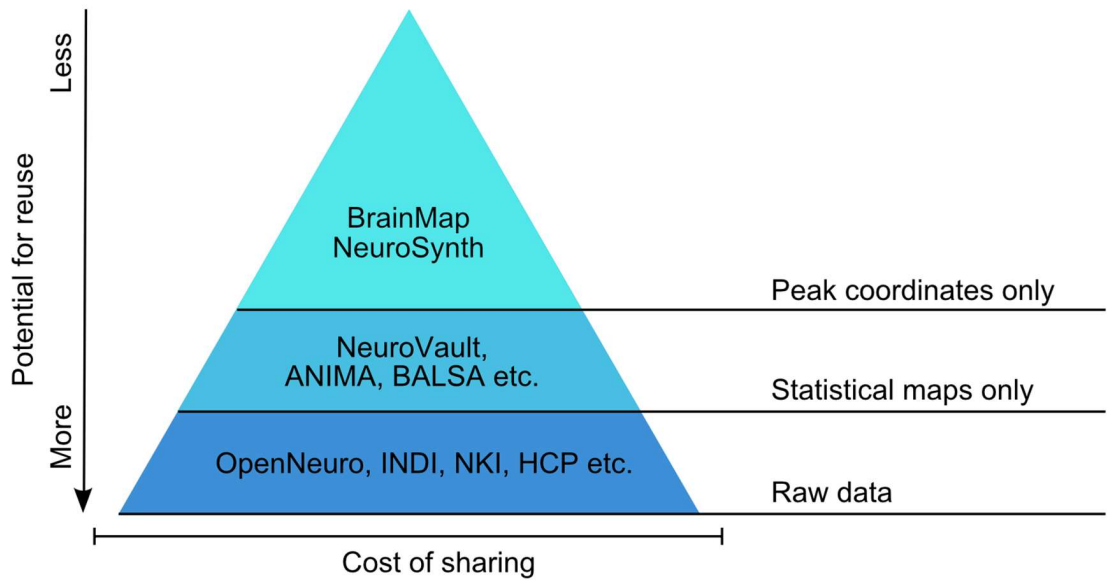

**Figure 1. Common levels of data sharing in neuroimaging.** Data sharing varies in the cost (both monetary and human time) required prepare and share the data, which is directly related to its potential for reuse. ANIMA: Archive of Neuroimaging Meta-analyses (20), BALSA: Brain Analysis Library of Spatial maps and Atlases , INDI: International Neuroimaging Data-Sharing Initiative (21), NKI: Nathan Kline Institute-Rockland Study (22), HCP: Human Connectome Project (23).

It is also worth noting that although "big data" is almost exclusively interpreted in terms of large numbers of individuals, the size of a dataset can actually vary along three important axes. The first, which we call "width", refers to the number of individuals participating in the study; this is the usual definition of "big data" in the context of neuroimaging. The second, which we call "breadth", refers to the number of phenotypes measured in the study. The third, which we call "depth", refers to the amount of measurement performed on each individual. Concerns about the statistical power of neuroimaging studies (1, 24) have led to a general increase in the width of datasets in the imaging literature. However, recent demonstrations of the value of broad and deep phenotyping (25–29) have led to increased interest in those versions of "big data" as well.

Within the neuroimaging community, there is a large set of data sharing initiatives that span the range from raw data to highly processed results. Individual researchers share small to medium size raw datasets (1-100 participants) using dedicated platforms such as OpenNeuro or NITRC. In other cases, multiple labs have joined forces to build consortia that share larger datasets (1,000-10,000 participants). This approach allows for spreading the financial burden of data collection across multiple entities and leads to diverse datasets (e.g. multiple scanners and sites) that allow better estimation of generalizability. Good examples of such initiatives are 1000 Functional Connectomes (14) and Consortium for Reliability and Reproducibility (30). Most consortia provide data access to all interested parties (pending adherence to user agreements), though some opt to only share data with their own members. While this approach may incentivize more labs to contribute data to the project in order to obtain access to the consortium data, it limits reusability of the data by other skilled researchers who do not have data to deposit.

Although sharing of Individual datasets and consortia have made a substantial amount of data accessible, the most prominent initiatives have involved large-scale prospective data acquisition and sharing. Those centrally funded initiatives involve planning data acquisition explicitly to maximize the potential for reuse, and often yield large sample sizes (up to 100,000 participants). Prospective planning allows for broader phenotyping meant to accommodate the needs of the entire community of scientists. The most successful such effort to date has been the Human Connectome Project (HCP)(23), which collected extensive imaging and other phenotypic data from 1,200 individuals during the period from 2009 to 2014. The data were shared prospectively in batches, and the complete dataset was released in 2017. Researchers can obtain the data through a data use agreement, with different types of data available through different levels of approval. The NIH has subsequently funded a number of extensions to the HCP, including ones focused on several aspects of lifespan development and a number of psychiatric and neurological diseases. The HCP data have had remarkable impact, leading to more than 600 publications (Jennifer Elam, personal communication). Importantly, this project has been an example of a data sharing "win-win", with the project investigators acheiving a number of high profile papers while still making the entire dataset publicly available with no restrictions on publication or requirements for coauthorship.

Other more recent initiatives will generate even larger datasets. The Adolescent Brain and Child Development (ABCD) study plans to image more than 10,000 children longitudinally for ten years, in order characterize brain development and its relation to mental illness and drug abuse(31). An even more ambitious effort is the UK Biobank, which plans to scan 100,000 participants and then follow them as they enter the age range for onset of common disorders of aging(32). Large scale initiatives are important because they provide large, carefully planned resources and homogeneous data, and are generally acquired using epidemiological approaches that result in more diverse and systematically acquired samples. The homogeneity of the data is potentially advantageous due to less variance in the data arising from data acquisition protocols and equipment. However, results generated using homogenous data may not generalize to data acquired using slightly different protocols or hardware. Additionally there is a potential concern with large studies that since they are very expensive, they tend to become flagship projects for funding agencies and risk overshadowing smaller initiatives and contributing to a lack of diversification of research investments. By focusing heavily on width, they also have the potential to crowd out other studies that could focus more heavily on depth or breadth.

### Low-cost approaches to data sharing and aggregation

Given the many diverse ways that data have been shared and the different varieties of data, there is a need for repositories and services that aggregate data. The simplest and most common form of shared data in neuroimaging is usually embedded in the manuscript itself, in the form of tables that list coordinates of the peaks of activation in a standard stereotactic format. Without aggregation this data would be difficult to use and require extensive literature reviews. To help with this, two projects have provided aggregation services. Brainmap (33) has generated a database of manually extracted coordinate data labelled according to the cognitive and behavioral features of the study. However, this publicly-funded resource has been unfortunately copyrighted, and gaining full access to it requires signing an extensive collaborative use agreement with the Principal Investigator that precludes open sharing of the data and requires coauthorship. A different approach, both in terms of data access and data aggregation, has been pursued by the Neurosynth project (34). The aim is very similar to Brainmap - extract and aggregate peak coordinate data - but it is achieved via automated text parsing algorithms rather than manual annotation. The accuracy of manual extraction and annotation is thus traded for scalability leading to larger coverage of the field. The Neurosynth database is openly shared with an accessible web API, released under the Open Database License (ODbL).

## Community-driven standards for data sharing

Shared data are only useful if they are formatted and organized in such a way that other researchers can fruitfully work with them. The minimum requirement is the use of common file formats that are broadly readable. Within the neuroimaging community, a common file format for imaging data known as NIfTI (Neuroimaging Informatics Technology Initiative) has emerged over the last two decades, and is now supported by all major neuroimaging software packages. While the NIfTI format is designed to represent volumetric data, other common formats have been developed to represent surface-based (Gifti) and combined surface/volume data (Cifti). The availability of software libraries to read and write these formats in a variety of languages has also enhanced their use.

Whereas common file formats are necessary for effective data sharing, they are not sufficient. First, because an imaging dataset usually comprises many different types of images, it is necessary to know which files correspond to which types of data, for which individual, etc. Second, for complex neuroimaging datasets it is also necessary to represent higher-order information regarding the dataset, for example, demographics of the individuals studied, or details of the cognitive task paradigm used in the experiment. To address these issues, in 2015 a group of neuroimaging researchers (supported by the International Neuroinformatics Coordinating Facility) convened to develop a new data organization standard for brain imaging data, which was called the Brain Imaging Data Structure (BIDS)(35). A draft standard was released for comment in September 2015, and version 1.0.0 of the standard was released in 2016.

The BIDS standard is an example of a Data Container specification (see Sidebar) and has two components. First, it provides a standard for file naming and directory organization, with file/directory naming templates that attempt to accommodate the large majority of use cases in neuroimaging research. The use of human-readable file organization reflects the fact that most research labs in neuroimaging use flat-file data storage rather than database storage; although other data organization schemes might be more powerful, the requirement for easy adoption by researchers led to the use of an approach that closely approximates common practices in the field.  The second component of the BIDS standard is a scheme for the organization of metadata.  These include metadata regarding data acquisition, cognitive paradigm and stimuli , as well as metadata regarding higher-order aspects of the dataset (such as authors and licensing). BIDS uses a common metadata format (JSON) to maximize the ease of writing compatible software. Each metadata term is defined in the specification and when possible linked to existing ontologies (such as DICOM).

### Validator-driven standards development

Development and implementation of a new data standard is a challenging task. Understanding the needs of all parties involved is crucial, and consequences of mistakes or inefficiencies of the proposed standard can cause issues for many years to follow, since necessary fixes can compromise backward compatibility. The BIDS community has adopted a *validator-driven* approach to standards development, wherein the standard is formally defined in terms of a validation tool that can check any dataset for adherence to the standard.

The BIDS Validator (see link in Resources) is a software package implemented in JavaScript that checks any dataset for standard compliance and issues a report outlining errors (missing required elements) or warnings (e.g. missing recommended elements or potential errors). Because it only examines image headers rather than loading entire images, it can process even very large datasets in a matter of seconds.  The validator plays three important roles in the BIDS ecosystem.  First, it has driven the standard developers to more formally specify the standard, and has highlighted important missing elements or edge cases. For example, writing JSON Schemas for the validator requires defining data types for dictionary fields: Is the field "Authors" a single string or a list of strings? Second, it has served as an important tool for users who are converting data into the standard, providing them with immediate feedback.  Third, the implementation in JavaScript has enabled the use of the BIDS Validator as a client-side element in the automated sharing of data via the OpenNeuro project, ensuring that any shared datasets meet the standard and preventing the need to upload large but possibly invalid datasets.

# Computational advances

In concert with the availability of increasingly rich neuroimaging datasets, analysis pipelines have also become increasingly complex. Reproducible use of such complex workflows requires a well-engineered software stack. Here we highlight the ways in which rapid advances in data-science software libraries can be elegantly combined with neuroimaging libraries to achieve reproducible and transparent analysis of large neuroimaging datasets.

## A consistent ecosystem is emerging based on the Python language

Neuroimaging analysis is both a programming and a modeling exercise. As a programming exercise, it calls for solid development practices in a language with modern features such as object-oriented programming, parallel computing, and package management. As a scientific modeling exercise, it requires rapid experimentation, simple and readable code, and interactive visualizations integrated with the analysis code. The Python language has emerged as a strong basis for tools that fulfill these needs, in neuroimaging as well in many other fields of science. General usage of Python is rapidly growing (see Figure 2). In 2018, Python is ranked as the third most used language worldwide[1], while MATLAB ranks 13th. As it is a widely-used general-purpose language used in many commercial settings, Python has many strong features including static code analysis to detect bugs, asynchronous computing, and an almost seamless interface to C, C++, or FORTRAN libraries. At the same time it is also a high-level language that can be used interactively, does not require variable typing or memory management, and hides a substantial amount of complexity from non-expert users. The marriage of both aspects has enabled expert developers to build scientific libraries that are easy to use for scientists without deep programming expertise.

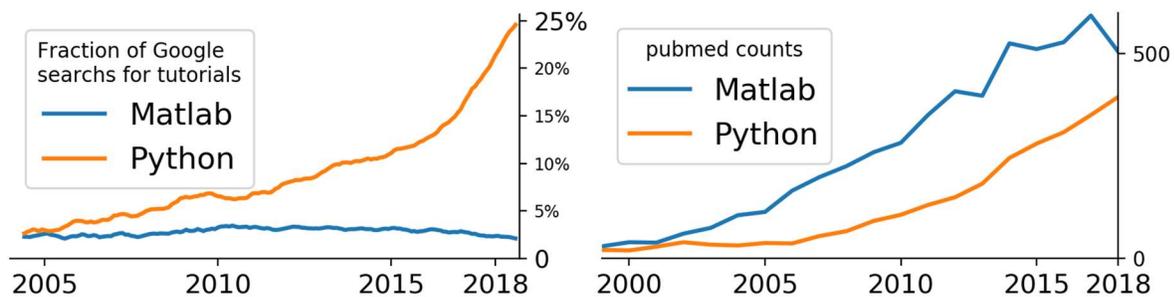

**Figure 2**: Usage of the Matlab and Python programming language; left: general usage (data: pypl), right: usage in biomedical research.

Scientific computing in Python revolves around the *numpy* library, which provides numerical arrays with a flexible memory layout and rich operations (36). This well-specified memory layout is crucial to exchange data across the stack without memory copies, including when calling external libraries written in C or FORTRAN. It has fostered the development of libraries exposing rich numerical-analysis algorithms in universal and easy-to-use interfaces. A central and prototypical example is the scipy library (37), which encompases many classic algorithms in applied mathematics including signal processing, optimization, Fourier transforms, linear algebra, interpolation, and more.

Another reason for the growth of Python in science is that it is an open-source stack that is strongly community-driven. Multiple communities, as varied as web developers, system administrators, or academic researchers, have built the Python ecosystem by contributing code and guidance, and the use of Python in large scale commercial enterprises such as Youtube, Instagram, Spotify, Industrial Light and Magic, and Redhat Linux provides further incentive to maintain and extend the language. This diversity of actors and interests also ensures long-term sustainability. The open-source model is well aligned with the objectives of reproducible science, since the code is available and can be audited by experts and there is an open tracking of issues and changes to the code over time. The fact that the software is distributed under a permissive license greatly facilitates academic usage, teaching, and large-scale deployments in cloud or cluster environments.

---

[1] https://www.tiobe.com/tiobe-index/

The Jupyter notebook has garnered particular interest within the open science community, as it provides easy interaction with code and execution on distant computers (38). It unites in a single interactive document, notes, blocks of code, their results, and interactive visualization (as in Figure 3). Such an environment lowers the barrier of entry to analyzing and understanding data. To produce reusable, library-level, code while interacting with data, there are plugins to integrate Jupyter functionality in advanced code editors such as atom, VScode, or Pycharm.

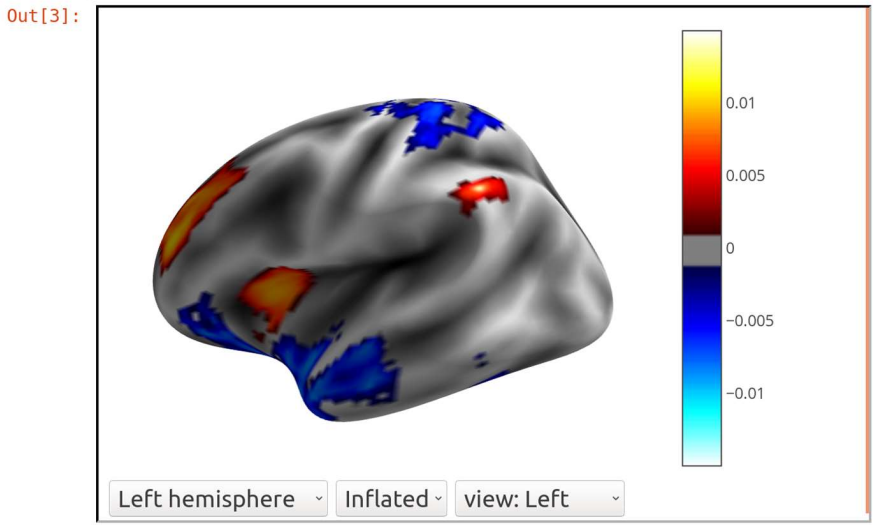

**Figure 3**: A Jupyter notebook, running an independent component analysis of rest-fMRI with nilearn and visualizing the results.

Unlike code written to analyze a dataset, library code must geared toward well-specified functionality that can be reliably reused. Good libraries require the use of software-engineering best practices (39). In particular, systematic automated testing helps ensure the validity of the options and the numerical stability of each function. Documentation and examples also need testing and maintenance, and an industry-grade language like Python comes with many tools to facilitate these tasks. As a result, the Python scientific-computing ecosystem has grown to provide robust tools to tackle large datasets, in particular in neuroimaging.

## Statistical learning packages enable more powerful analysis

Advances in multivariate statistics combined with computational thinking have fueled a revolution in data processing, combining methods from statistics and computer science into the new field known as "data science". The core scientific progress happens in the field of machine learning, which combines statistics and algorithms to fit models that are tuned for *prediction*, unlike conventional statistics which is more focused on testing model parameters. This paradigm shift in data modeling is important as it enables the use of more complex models, for which statistical control of the parameters would be very different. In machine learning, a model is useful and valid if it accurately predicts unseen data. Machine learning has opened new alleys in extracting information from texts, images, genomes, etc (40), with applications ranging from spam detection to medical diagnosis (41).

Many scientific data-processing problems can be reformulated with the help of predictive models, including in psychology and brain imaging (42, 43). Various applications of brain imaging draw different benefits from machine learning (43). In cognitive neuroscience, models that generalize explicitly to new data or new conditions provide the basis for establishing generalizable associations between mind and brain (44). Supervised machine learning can be used to link brain activity to a corresponding mental state, as created in a psychological experiment, allowing the ability to *decode* psychological states from neuroimaging data (45). In clinical applications, predicting individual traits from brain images can provide potential biomarkers for psychiatric or neurological disorders (46); in this case, prediction is typically performed across individuals, in a population study (47).

All these approaches have strongly benefitted from progress in data sharing, as it has enabled them to learn more general markers from more diverse data. For instance, in biomarker development, pooling sites can to lead to markers that are robust to site variance (48). For decoding mental processes, probing various types of pain –physical and social– reveals a clear signature of physical pain (49), an analysis enabled by pooling across multiple studies. Indeed, multiple studies bring different paradigms that investigate a variety of mental processes with different psychological manipulations. Decoding mental processes allows characterization of the functions of particular brain structures that support such prediction: decoding across many studies shows that activity in these regions indeed does imply the corresponding mental process (50). At the level of a single study, this conclusion would be an invalid *reverse inference*, as the data only shows that the activity is a consequence of the behavior triggered in the experiment, and not the cause (51). Aggregating many studies is thus needed to draw general conclusions on links between brain and cognition, and progress in data sharing has made this aggregation much easier.

In computational anatomy settings, predictive models have been used to segment particular features from images, *eg* predicting the presence of lesions. Here larger databases of brain images have enabled training of richer machine learning models that lead to improved segmentation of brain structures (52, 53).

Predicting from brain images raises specific challenges to machine learning methods, as the data –3D or 4D images– are very complex, with a large number of features (in the hundreds of thousands), whereas the number of observations is usually quite small in comparison. For functional imaging, data accumulation is often easier outside of controlled psychological manipulations, such as resting-state fMRI (14). With such recordings, the time series themselves do not contain any consistent features as they would in a controlled task; machine learning is instead performed on a *functional connectome*, built from the correlation across features (54), to distinguish brain states or individual traits.

Machine learning also provides tools to characterize structure in data without the presence of explicit labels to predict, known as *unsupervised learning*. For instance, applied to many brain images without labels, unsupervised learning can build brain parcellations, *eg* by clustering voxels (55), or extract brain networks from resting-state data using matrix factorization techniques such as independent component analysis (56). A significant benefit of unsupervised learning, compared to supervised learning, is that it can leverage data without extra information on mental or clinical state, which are easier to collect and to share than more deeply annotated datasets. With large collections of resting-state acquisition, unsupervised learning, for instance independent component analysis (ICA) – see Figure 3– extracts brain networks or regions that form a good basis for supervised learning on resting-state (48) or on task data (57).

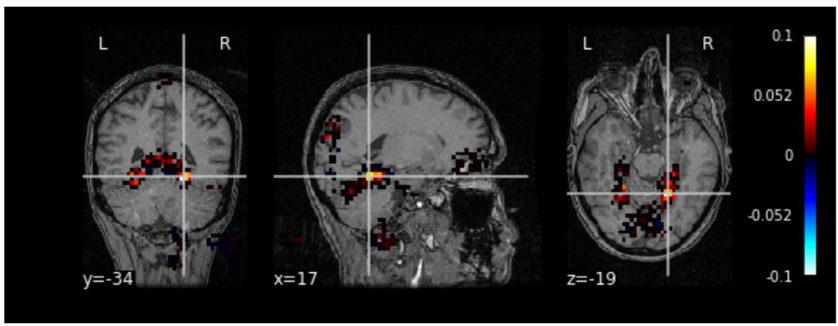

```
In [1]:  # Load behavioral data
         import pandas as pd
         behavioral = pd.read_csv('/home/varoquau/nilearn_data/haxby2001/subj1/labels.txt', sep=" ")

         # Restrict to face and house conditions
         conditions = behavioral['labels']
         condition_mask = conditions.isin(['face', 'house'])

         # Split data into train and test samples, using the chunks
         condition_mask_train = (condition_mask) & (behavioral['chunks'] <= 6)
         condition_mask_test = (condition_mask) & (behavioral['chunks'] > 6)
         # Because the data is in one single large 4D image, we need to use index_img to do the split
         from nilearn.image import index_img
         train_imgs = index_img('/home/varoquau/nilearn_data/haxby2001/subj1/bold.nii.gz', condition_mas
         test_imgs = index_img('/home/varoquau/nilearn_data/haxby2001/subj1/bold.nii.gz', condition_mask
         train_conditions = conditions[condition_mask_train]
         test_conditions = conditions[condition_mask_test]
```

Fit a decoding on test data

```
In [2]:  from nilearn.decoding import SpaceNetClassifier
         decoder = SpaceNetClassifier(verbose=0)
         decoder.fit(train_imgs, train_conditions)
```

...

Plot the decoding regions

```
In [3]:  %matplotlib inline
         from nilearn import plotting
         display = plotting.plot_stat_map(decoder.coef_img_, dim=-1,
                                          bg_img='/home/varoquau/nilearn_data/haxby2001/subj1/anat.nii.g
```

Measure the accuracy of predicting behavior on unseen data

```
In [4]:  predicted_conditions = decoder.predict(test_imgs)
         accuracy = (predicted_conditions == test_conditions).mean() * 100.
         print("Classification accuracy : %g%%" % accuracy)

         Classification accuracy : 91.1111%
```

**Figure 4:** A complete decoding analysis with nilearn: learning to discriminate whether a subject is seeing faces or places from brain activity.

The availability of high-quality machine learning libraries has fostered their adoption in neuroimaging. The first widespread adoption came from libsvm (58), via its MATLAB binding. Within the Python ecosystem, the scikit-learn library (59) provides a very versatile tool with many statistical models and utilities, and has become heavily used for machine learning analyses of neuroimaging data.

Beyond the classic models, there is active methods research to develop machine-learning models tailored to the specifics of brain imaging data. Accounting for the properties of the data, such as its spatial structure, can bring benefits to statistical analysis. Yet, for reproducible research it is a double-edged sword as the multiplication of methods increases the analytic flexibility of researchers, widening the "garden of forking paths" (60). For this reason, careful validation of methods is paramount: as new methods are developed, standard pipelines should be reassessed with the goal of selecting only a small number of recommended options. Current practices in neuroimaging unfortunately span variations in analytic choices that are arguably too broad (61). As the application of machine learning to brain imaging data is comparatively more recent, it is still often performed via custom-written code for which reproducibility is harder to ensure.

## Neuroimaging-specific libraries for the last mile

Neuroimaging-specific libraries are important for reproducibility of results of neuroimaging analysis, but also to make such analysis easier. Analyzing brain images calls for a complex pipeline composed of many steps. The pipeline typically starts with a set of spatial or computational neuroanatomy steps that strive to identify anatomical structures and align them across images or across subjects. It is then followed by statistical modeling steps that characterize the link between a particular psychological or clinical effect and the data. There are software challenges for steps of the pipeline, ranging from assembling many diverse tools to enabling and recording analysis choices and tracking provenance of the resulting derivative data. Recent advances in the neuroimaging software landscape has brought tools that make such complex analyses easier and more reproducible.

To tackle the many steps and software components that can be combined in an end-to-end analysis pipeline, Nipype (62) provides Python interfaces to all of the major command-line tools used by neuroimaging research. As a result, it enables users to leverage classic neuroimaging software packages, such as Freesurfer (63), FSL (64) or ANTS (65), that have traditionally been combined via custom shell scripts. With nipype, users write these pipelines in the Python language, which is much more structured than a shell script. The resulting code is more universal when combining tools. Type checking of the input and output of each step leads to better error management. In addition, it makes it easy to vary parameters of pipeline. Nipype also comes with a dataflow engine, an explicit description of how data and parameters flow through a processing pipeline. This enables the speedup of re-execution of a pipeline by computing only steps that changed, and also supports efficiently distribution of computations across a computing cluster, managing data dependencies between the different steps and consequently minimizing data transfer across the cluster. This scalability is important to tackle increasingly large brain imaging datasets.

The neuroimaging preprocessing pipelines are used to align brain images to a common reference across subjects and to output features or maps describing specific anatomical structures. These can then be readily used in a statistical modeling or machine learning analyses. Nilearn (66) is a Python library that facilitates typical applications of machine-learning techniques to brain imaging data, combining models from scikit-learn with neuroimaging-specific code. An important aspect of nilearn is that it provides many tools that go from the data given in a brain-imaging specific representation, such as a Nifti file, to a more abstract data-matrix, the typical input of a machine-learning model, where features of the data are numerical columns and observations are rows. As can be seen from Figure 4, an end-to-end machine-learning workflow written with nilearn is fairly short and expressive, starting directly from the neuroimaging data and finishing with a visualization adapted to the needs of neuroimaging experts.

Generic machine-learning models do not capture all properties of the data, such as the fact that features are on a three-dimensional grid or a two-dimensional surface; nilearn thus also provides models that are specific to neuroimaging data. Indeed, capturing the spatial structure is often important, for instance in decoding, to find the regions that predict behavior (67, 68), or when clustering brain-activity time series, to extract regions of coherent activity in the absence of task (69). Having a library of models tailored to brain imaging modalities also helps by providing useful defaults and a combination of general-purpose machine-learning algorithms that properly tackle tasks of interest to the scientist. Removing from users the burden of assembling signal-processing steps and building a complex pipeline makes them more productive and makes the code easier to read, edit, and reproduce. Nilearn however remains a very versatile tool, in that it does not make all the choices for the user. Indeed, it can be employed to solve many different machine learning tasks –from decoding or encoding cognition, to predictive modeling based on anatomical or resting-state data. Narrowing the usage patterns and applications gives even tighter code, such as that developed in Pymvpa (70) which focuses on a smaller number of decoding methods.

Most brain images are volumetric images, acquired via magnetic resonance (MR) or positron emission tomography (PET) imaging. However, other imaging modalities involving electrophysiological measurements are of great interest to neuroscience and psychology, in particular electroencephalography (EEG) and magnetoencephalography (MEG). MNE-Python (71) is a Python toolbox focused on analysing these data. It also provides connections to scikit-learn machine learning models, for instance used in decoding applications. The Python ecosystem (72) for brain imaging is strong and diverse, with many libraries of various size and focus, such as –in addition to these mentioned previously– dipy (73) for processing of diffusion-weighted MRI data, or nibabel (74), which provides tools to read and operate on many of the file formats used in neuroimaging.

## The role of visualization in reproducibility

Producing figures in an important aspect in the full process of going from data to a scientific publication. Figures that are didactic and visually engaging improve the impact of publications. However, for transparency and reproducibility, it is essential that figures can be linked directly to the data and processing steps used to create them. Programmatically generating beautiful figures from data is difficult, because there are many different ways of representing data, because it requires a careful attention to details, and because choices are often more easily specified interactively. Visualisation is much less often discussed in the context of reproducibility than data processing and statistical analysis. However, powerful libraries that can generate rich and meaningful figures with simple code from data also contribute significantly to reproducibility. Within the Python ecosystem, one general-purpose visualization toolbox of note is the Seaborn package (see Resources).

Visualization is also important during the exploratory step of data analysis, for instance to understand the data, perform quality assurance, or debug an analysis pipeline. Interactive visualization is a strong benefit for such tasks. Yet interactivity can quickly come at the cost of reproducibility. Hence visualization tools navigate these competing goals, exploring trade offs or pushing scripting as far as possible. Visualization libraries for brain imaging make it easy to represent data in a way that is meaningful to neuroscientists, for instance by relating it to neuroanatomical landmarks. This task is challenging because the data are three-dimensional, sometimes four-dimensional. Yet, as a research field, neuroimaging has developed standard displays that summarize well the information. For reproducibility, they can be used in data analysis scripts, for instance with nilearn (see Figures 3 and 4).

## Software engineering for reproducible science

Being able to reproduce the findings of an academic paper increases faith in the presented results. However, despite growing availability of data and a mature ecosystem of computational libraries, numerical reproducibility is still hard to achieve. Existing libraries and tools can be combined in many different ways that are not always captured by the methods section of a manuscript. Furthermore, results can differ drastically depending on platform, software library version, and even order of compilation (75). Numerical reproducibility requires not only access to data and code, but also a snapshot of the environment used for the analysis (capturing all of the dependencies and their configuration). This feat can be achieved via software containers - a lightweight successor to virtual machines that capture the entire software stack above the operating system kernel. Two most popular implementations of this technology are Docker (well suited for cloud deployment and desktop development) and Singularity (preferred to Docker in multi tenant environments such as HPCs) (76). Since they both capitalize on the Linux kernel, the building of reproducible container images benefits from repositories of Linux neuroimaging software such as Neurodebian (77).

Software containers can be extremely useful for capturing the environment for a given project for the purpose of executing the same analysis on another machine (i.e. prototype on a desktop - run on an HPC system). They are also helpful in the context of longitudinal studies with ongoing analyses, and for switching between projects. Software containers can also be a useful mean to distribute analysis pipelines with complex binary dependencies. This idea has been the key feature of BIDS Apps - portable neuroimaging pipelines with a common command line interface and the ability to parse BIDS datasets as an input (78). BIDS Apps consist of a growing collection of existing neuroimaging pipelines conformed to the standard of inputs and encapsulated in a container. This not only helps with installation issues, but also increases reproducibility of analyses since each container image has a unique version that can be used to rerun the analysis. All BIDS Apps also use automated testing via continuous integration systems, to improve software quality and prevent regressions.

### The glass box philosophy of tool development

Ease of installation and use can unfortunately come at a cost: Lack of understanding of the minutiae of a tool can lead to misuse or misinterpretation of the result by users. This trade-off can be to some extent mitigated by adapting the *glass box* philosophy of software development (first introduced in context of the FMRIPREP tool - (79)). This approach follows three rules: I) writing thorough didactic documentation that lets user understand how the tool works, II) providing visual reports explaining the results and intermediate steps for each invocation of the tool, and III) assisting users with accurate dissemination of the methods encapsulated in the tool by providing boilerplate text ready to use in a methods section of a paper. Glass box does not completely remove the risk of misuse, but reduces the cost of automation while still democratizing access to powerful and well-engineered but complex methods.

Lack of easy access to reproducible tools and workflows, especially those that are computationally expensive, can reduce adoption. One way to improve this is to provide the ability to perform reproducible analyses accessible via the browser, an approach known as "Science as a service". Platforms such as OpenNeuro and commercial counterparts such as Kaggle or CodeOcean allow users to share and version their data as well as run versioned and reproducible pipelines. OpenNeuro, for example, heavily relies on the BIDS standard and allows user to run BIDS Apps. Each execution is performed on an immutable snapshot (version) of an input dataset using a specific version of requested BIDS App. Data versioning and software containers thus provide analysis reproducibility neatly packaged in a web interface. There are many challenges to running Science as a Service platforms, particularly with regard to sustainability given the ongoing cost of computing and data storage, but it is so far the the most user-friendly model for computational reproducibility.

### Archival reproducibility does not guarantee validity

Controlling the environment and archiving data and processing operations ensures that results are reproducible, but not that they are correct. A computational pipeline can be considered as correct only if it performs the operations that it claims it should. However, validity is not always easy to define or control. In the case of data analysis, correct control of statistics, such as p-values or false discovery rate, is even harder to ensure, as it depends upon assumptions about the data; further, a pipeline composed of many steps is difficult to study mathematically. In neuroimaging, studies using resting fMRI as "null" data have shown that invalid assumptions about spatial autocorrelation in a popular software have led to inflated error rates (80).

Statistical assumptions of data-processing pipelines must be checked empirically, validating them in multiple ways on multiple datasets. Extra efforts are needed to ensure that pipelines provide the same results on different platforms, as small numerical errors –such as differences in floating-point representations– can lead to substantially different end results (81). Beyond statistical assumptions or numerical instabilities, logical errors –bugs– where the analysis does not behave as expected, can hide in the code (82). When these errors do not lead to visible failures of the pipeline, they can lead researchers to publish results obtained with incorrect analysis, or simple analyses that deviate heavily from the methods reported in the publication. Of particular concern is that researchers are less likely to detect bugs that produce results that confirm their hypotheses compared to those that disconfirm them.

To check for potential errors, a publication must build upon open code that can be analysed a posteriori, but analysing and understanding code is difficult. The respected software developer Brian Kernighan wrote: "Everyone knows that debugging is twice as hard as writing a program in the first place. So if you're as clever as you can be when you write it, how will you ever debug it?" (83 chap 2). Code must be curated and maintained, to ensure that it is correct and robust. For these reasons, it is important that analysis code be turned into libraries, sharing efforts across many developers. Indeed, "many eyes make all bugs shallow" (84). Libraries also facilitate formal testing of the code, which is crucial to ensure that it is correct.

The more a publication relies on standard libraries rather than custom code, the more understable it is and the less likely it is to contain major bugs. Archiving complex and custom code, e.g. in a software container, enables reproducibility, though the long-term sustainability of any particular container system is always uncertain. In addition, although containers provide some degree of transparency, their complexity can make this somewhat challenging (e.g. through inclusion of binaries without source code). The use of modified library versions within a container can also result in a processing stream that diverges from community standards. Archival reproducibility is necessary but not sufficient for a healthy scientific process; While reproducibility of analyses builds the memory of science, reusability builds its future.

# Conclusion

Ensuring the quality of scientific research is an ongoing battle, which is made increasingly challenging by the availability of large datasets and complex analysis workflows. Reproducibility requires openness and transparency, which has been greatly enhanced in the field of neuroimaging by open data sharing and the use of open source software. The growth of software complexity requires increasing sophistication in software engineering methods, which motivates the development of high-quality software libraries and analysis platforms that follow a "glass box" philosophy. The success of an open science ecosystem within the field of neuroimaging provides a guiding example for researchers in other fields.

# Future Issues list:

- Long-term sustainability is a continuing challenge for community data sharing projects, given the inability of funding agencies to guarantee long-term support. How can we ensure that data will remain accessible in the long term?

- What is the best model for analyzing petascale neuroimaging datasets that are challenging to copy between sites?
- Researchers are driven by institutional incentives (for publications, grants, and academic credit) to develop multiple databases within the same sphere of research. How can incentives be realigned to promote the development of collaborative databases?
- The expanding landscape of data sharing across multiple projects drives a need for better ways for searching through available data. The recent development of Google Datasets provides a potential resource, but further work is needed to identify the best approach for indexing and searching the range of openly available datasets.

## Summary Points List:

- Open data and analysis software are essential for reproducibility in data analysis
- Data sharing has grown from the ground up within the neuroimaging community, with large databases of open data now available.
- The Brain Imaging Data Structure has provided a common language in which to describe and organize a broad range of neuroimaging datasets.
- The Python ecosystem for neuroimaging provides a set of open source tools for reproducible and transparent analysis of neuroimaging data.
-

## Acknowledgments:

The work described here was supported by National Institutes of Health (R24MH114705 and R24MH117179), National Science Foundation (IIS-1760950), and the Laura and John Arnold Foundation.

**Acronyms and Definitions list (glossary):**

BIDS: Brain Imaging Data Structure - A data container standard for neuroimaging datasets
fMRI: functional magnetic resonance imaging - The primary method for functional imaging of the human brain
PET: Positron emission tomography - A method for metabolic imaging of the brain
MEG: Magnetoencephalography - A method for imaging of electrical signals in the brain via measurement of their electromagnetic signatures
EEG: Electroencephalography - A method for imaging of electrical signals in the brain via recordings from electrodes on the scalp
INDI: International Neuroimaging Data-sharing Initiative - A data sharing initiative that organized the 1000 Functional Connectomes Project
HCP: Human Connectome Project - A major NIH-funded effort to generate a shared dataset intended to map the human connectome.

NIfTI: Neuroimaging Informatics Technology Initiative: A standard format for storage of neuroimaging data
JSON: JavaScript Object Notation - A standard format for structured data storage, used in the BIDS standard
DICOM: Digital Imaging and Communications in Medicine - A standard format for the storage of medical imaging data
ICA: independent component analysis - A matrix factorization technique commonly used in the neuroimaging field.
NITRC - Neuroimaging Informatics Tools and Resources Clearinghouse - A portal for software and data resources in neuroimaging

**Related Resources list:**

- [http://scikit-learn.org/](http://scikit-learn.org/): The scikit-learn project for machine learning
- [http://bids.neuroimaging.io](http://bids.neuroimaging.io) - The BIDS standard for neuroimaging data containers
- [https://incf.github.io/bids-validator/](https://incf.github.io/bids-validator/): The BIDS validator, which performs client-side validation using the BIDS standard
- [https://bids-apps.neuroimaging.io/](https://bids-apps.neuroimaging.io/): The BIDS-Apps specification for containerized BIDS-aware software applications
- [https://nilearn.github.io/](https://nilearn.github.io/): The nilearn project for machine learning analysis in Python
- [http://openneuro.org](http://openneuro.org): The OpenNeuro data sharing and analysis project
- [https://www.nitrc.org/](https://www.nitrc.org/): The Neuroimaging Informatics Tools and Resources Clearinghouse

**Sidebar**

> **Data Containers (sidebar)**
> Data container is a data science concept that describes a dataset organization and annotation that makes the dataset maximally reusable without the need for additional information. In other words, all of the data and metadata necessary to use the data are contained in the same structure. This property makes data containers an excellent data exchange format, both in the sense of communication between two databases as well as traditional data sharing between individual researchers. EEG Study Schema (ESS - (85)) and BIDS are good examples of data containers. BIDS serves as a data exchange format in OpenNeuro.org (all incoming datasets need to be BIDS) as well as SchizConnect (queries across multiple data sources return a single data container in BIDS format).

Literature Cited

1. Button KS, Ioannidis JPA, Mokrysz C, et al. 2013. Power failure: why small sample size undermines the reliability of neuroscience. *Nat. Rev. Neurosci.* 14(5):365–76
2. Open Science Collaboration. 2015. Estimating the reproducibility of psychological science. *Science*. 349(6251):aac4716
3. Errington TM, Iorns E, Gunn W, et al. 2014. Science forum: An open investigation of the reproducibility of cancer biology research. *Elife*. 3:e04333
4. Herndon T, Ash M, Pollin R. 2014. Does high public debt consistently stifle economic growth? A critique of Reinhart and Rogoff. *Cambridge J. Econ.* 38(2):257–79
5. Christensen GS, Miguel E. 2016. *Transparency, Reproducibility, and the Credibility of Economics Research*. Work. Pap.
6. Poldrack RA, Farah MJ. 2015. Progress and challenges in probing the human brain. *Nature*. 526(7573):371–79
7. Peng RD. 2011. Reproducible research in computational science. *Science*. 334(6060):1226–27
8. Goodman SN, Fanelli D, Ioannidis JPA. 2016. What does research reproducibility mean? *Sci. Transl. Med.* 8(341):341ps12
9. Patil P, Peng RD, Leek J. 2016. *A statistical definition for reproducibility and replicability*. Work. Pap.
10. Longo DL, Drazen JM. 2016. Data Sharing. *N. Engl. J. Med.* 374(3):276–77
11. Drazen JM. 2016. Data Sharing and the Journal. *N. Engl. J. Med.* 374(19):e24
12. Greene CS, Garmire LX, Gilbert JA, et al. 2017. Celebrating parasites. *Nat. Genet.* 49(4):483–84
13. Van Horn JD, Gazzaniga MS. 2013. Why share data? Lessons learned from the fMRIDC. *Neuroimage*. 82:677–82
14. Biswal BB, Mennes M, Zuo X-N, et al. 2010. Toward discovery science of human brain function. *Proc. Natl. Acad. Sci. U. S. A.* 107(10):4734–39

15. Gorgolewski KJ, Wheeler K, Halchenko YO, et al. 2015. The impact of shared data in neuroimaging: the case of OpenfMRI.org

16. Milham MP, Craddock RC, Son JJ, et al. 2018. Assessment of the impact of shared brain imaging data on the scientific literature. *Nat. Commun.* 9(1):2818

17. Grisham W, Brumberg JC, Gilbert T, et al. 2017. Teaching with Big Data: Report from the 2016 Society for Neuroscience Teaching Workshop. *J. Undergrad. Neurosci. Educ.* 16(1):A68–76

18. Saidi HIB. 2018. *Power Comparisons of the Rician and Gaussian Random Fields Tests for Detecting Signal from Functional Magnetic Resonance Images*. University of Northern Colorado

19. Ioannidis JPA. 2008. Why most discovered true associations are inflated. *Epidemiology*. 19(5):640–48

20. Reid AT, Bzdok D, Genon S, et al. 2016. ANIMA: A data-sharing initiative for neuroimaging meta-analyses. *Neuroimage*. 124(Pt B):1245–53

21. Mennes M, Biswal BB, Castellanos FX, et al. 2013. Making data sharing work: the FCP/INDI experience. *Neuroimage*. 82:683–91

22. Nooner KB, Colcombe SJ, Tobe RH, et al. 2012. The NKI-Rockland Sample: A Model for Accelerating the Pace of Discovery Science in Psychiatry. *Front. Neurosci.* 6:152

23. Van Essen DC, Smith SM, Barch DM, et al. 2013. The WU-Minn Human Connectome Project: an overview. *Neuroimage*. 80:62–79

24. Poldrack RA, Baker CI, Durnez J, et al. 2017. Scanning the horizon: towards transparent and reproducible neuroimaging research. *Nat. Rev. Neurosci.* 18(2):115–26

25. Braga RM, Buckner RL. 2017. Parallel Interdigitated Distributed Networks Within the Individual Estimated by Intrinsic Functional Connectivity. *Neuron*

26. Gordon EM, Laumann TO, Gilmore AW, et al. 2017. Precision Functional Mapping of Individual Human Brains. *Neuron*. 95(4):791–807.e7

27. Poldrack RA. 2017. Precision Neuroscience: Dense Sampling of Individual Brains. *Neuron*. 95(4):727–29

28. Pinho AL, Amadon A, Ruest T, et al. 2018. Individual Brain Charting, a high-resolution fMRI dataset for cognitive mapping. *Sci Data*. 5:180105

29. Poldrack RA, Laumann TO, Koyejo O, et al. 2015. Long-term neural and physiological phenotyping of a single human. *Nat. Commun.* 6:8885

30. Zuo X-N, Anderson JS, Bellec P, et al. 2014. An open science resource for establishing reliability and reproducibility in functional connectomics. *Sci Data*. 1:140049

31. Casey BJ, Cannonier T, Conley MI, et al. 2018. The Adolescent Brain Cognitive Development (ABCD) study: Imaging acquisition across 21 sites. *Dev. Cogn. Neurosci.* 32:43–54

32. Miller KL, Alfaro-Almagro F, Bangerter NK, et al. 2016. Multimodal population brain imaging in the UK Biobank prospective epidemiological study. *Nat. Neurosci.* 19(11):1523–36

33. Fox PT, Lancaster JL. 2002. Opinion: Mapping context and content: the BrainMap model. *Nat. Rev. Neurosci.* 3(4):319–21

34. Yarkoni T, Poldrack RA, Nichols TE, et al. 2011. Large-scale automated synthesis of human functional neuroimaging data. *Nat. Methods*. 8(8):665–70

35. Gorgolewski KJ, Auer T, Calhoun VD, et al. 2016. The brain imaging data structure, a format for organizing and describing outputs of neuroimaging experiments. *Sci Data*. 3:160044

36. Walt S van der, Colbert SC, Varoquaux G. 2011. The NumPy Array: A Structure for Efficient Numerical Computation. *Comput. Sci. Eng.* 13(2):22–30

37. Jones E, Oliphant T, Peterson P. 2001--. {SciPy}: Open source scientific tools for {Python}

38. Kluyver T, Ragan-Kelley B, Pérez F, et al. 2016. Jupyter Notebooks-a publishing format for reproducible computational workflows

39. Wilson G, Aruliah DA, Brown CT, et al. 2014. Best practices for scientific computing. *PLoS*

*Biol.* 12(1):e1001745

40. LeCun Y, Bengio Y, Hinton G. 2015. Deep learning. *Nature*. 521(7553):436–44

41. Jordan MI, Mitchell TM. 2015. Machine learning: Trends, perspectives, and prospects. *Science*. 349(6245):255–60

42. Yarkoni T, Westfall J. 2017. Choosing Prediction Over Explanation in Psychology: Lessons From Machine Learning. *Perspect. Psychol. Sci.* 12(6):1100–1122

43. Varoquaux G, Thirion B. 2014. How machine learning is shaping cognitive neuroimaging. *Gigascience*. 3:28

44. Varoquaux G, Poldrack R. 2018. Predictive models can overcome reductionism in cognitive neuroimaging

45. Pereira F, Mitchell T, Botvinick M. 2009. Machine learning classifiers and fMRI: a tutorial overview. *Neuroimage*

46. Woo C-W, Chang LJ, Lindquist MA, et al. 2017. Building better biomarkers: brain models in translational neuroimaging. *Nat. Neurosci.* 20(3):365–77

47. Klöppel S, Abdulkadir A, Jack CR Jr, et al. 2012. Diagnostic neuroimaging across diseases. *Neuroimage*. 61(2):457–63

48. Abraham A, Milham MP, Di Martino A, et al. 2017. Deriving reproducible biomarkers from multi-site resting-state data: An Autism-based example. *Neuroimage*. 147:736–45

49. Wager TD, Atlas LY, Lindquist MA, et al. 2013. An fMRI-based neurologic signature of physical pain. *N. Engl. J. Med.* 368(15):1388–97

50. Poldrack RA. 2011. Inferring mental states from neuroimaging data: from reverse inference to large-scale decoding. *Neuron*. 72(5):692–97

51. Poldrack RA. 2006. Can cognitive processes be inferred from neuroimaging data? *Trends Cogn. Sci.* 10(2):59–63

52. Menze BH, Jakab A, Bauer S. 2015. The multimodal brain tumor image segmentation benchmark (BRATS). *IEEE transactions on*

53. Shin H-C, Roth HR, Gao M, et al. 2016. Deep Convolutional Neural Networks for Computer-Aided Detection: CNN Architectures, Dataset Characteristics and Transfer Learning. *IEEE Trans. Med. Imaging*. 35(5):1285–98

54. Varoquaux G, Craddock RC. 2013. Learning and comparing functional connectomes across subjects. *Neuroimage*. 80:405–15

55. Thirion B, Varoquaux G, Dohmatob E, et al. 2014. Which fMRI clustering gives good brain parcellations? *Front. Neurosci.* 8:167

56. Kiviniemi V, Kantola J-H, Jauhiainen J, et al. 2003. Independent component analysis of nondeterministic fMRI signal sources. *Neuroimage*. 19(2 Pt 1):253–60

57. Mensch A, Mairal J, Bzdok D, et al. 2017. Learning Neural Representations of Human Cognition across Many fMRI Studies. In *Advances in Neural Information Processing Systems 30*, ed I Guyon, UV Luxburg, S Bengio, et al., pp. 5883–93. Curran Associates, Inc.

58. Chang C-C, Lin C-J. 2011. LIBSVM: A Library for Support Vector Machines. *ACM Trans. Intell. Syst. Technol.* 2(3):27:1–27:27

59. Pedregosa F, Varoquaux G, Gramfort A, et al. 2011. Scikit-learn: Machine Learning in Python. *J. Mach. Learn. Res.* 12(Oct):2825–30

60. Gelman A, Loken E. 2013. The garden of forking paths: Why multiple comparisons can be a problem, even when there is no "fishing expedition" or "p-hacking" and the research hypothesis was posited ahead of time. *Department of Statistics, Columbia University*

61. Carp J. 2012. On the plurality of (methodological) worlds: estimating the analytic flexibility of FMRI experiments. *Front. Neurosci.* 6:149

62. Gorgolewski K, Burns CD, Madison C, et al. 2011. Nipype: a flexible, lightweight and extensible neuroimaging data processing framework in python. *Front. Neuroinform.* 5:13

63. Fischl B. 2012. FreeSurfer. *Neuroimage*. 62(2):774–81

64. Smith SM, Jenkinson M, Woolrich MW, et al. 2004. Advances in functional and structural

MR image analysis and implementation as FSL. *Neuroimage*. 23 Suppl 1:S208–19

65. Avants BB, Tustison N, Song G. Advanced normalization tools (ANTS). *scil.dinf.usherbrooke.ca*

66. Abraham A, Pedregosa F, Eickenberg M, et al. 2014. Machine learning for neuroimaging with scikit-learn. *Front. Neuroinform.* 8:14

67. Michel V, Gramfort A, Varoquaux G, et al. 2011. Total variation regularization for fMRI-based prediction of behavior. *IEEE Trans. Med. Imaging*. 30(7):1328–40

68. Grosenick L, Klingenberg B, Katovich K, et al. 2013. Interpretable whole-brain prediction analysis with GraphNet. *Neuroimage*. 72:304–21

69. Craddock RC, James GA, Holtzheimer PE 3rd, et al. 2012. A whole brain fMRI atlas generated via spatially constrained spectral clustering. *Hum. Brain Mapp.* 33(8):1914–28

70. Hanke M, Halchenko YO, Sederberg PB, et al. 2009. PyMVPA: A python toolbox for multivariate pattern analysis of fMRI data. *Neuroinformatics*. 7(1):37–53

71. Gramfort A, Luessi M, Larson E, et al. 2014. MNE software for processing MEG and EEG data. *Neuroimage*. 86:446–60

72. Millman KJ, Brett M. 2007. Analysis of Functional Magnetic Resonance Imaging in Python. *Comput. Sci. Eng.* 9(3):52–55

73. Garyfallidis E, Brett M, Amirbekian B, et al. 2014. Dipy, a library for the analysis of diffusion MRI data. *Front. Neuroinform.* 8:8

74. Brett M, Hanke M, Cipollini B, et al. 2016. nibabel: 2.1. 0. *Zenodo*

75. Mackenzie-Graham AJ, Van Horn JD, Woods RP, et al. 2008. Provenance in neuroimaging. *Neuroimage*. 42(1):178–95

76. Kurtzer GM, Sochat V, Bauer MW. 2017. Singularity: Scientific containers for mobility of compute. *PLoS One*. 12(5):e0177459

77. Halchenko YO, Hanke M. 2012. Open is not enough. Let' s take the next step: An integrated, community-driven computing platform for neuroscience. *Front. Neuroinform.*

6(22):

78. Gorgolewski KJ, Alfaro-Almagro F, Auer T, et al. 2017. BIDS apps: Improving ease of use, accessibility, and reproducibility of neuroimaging data analysis methods. *PLoS Comput. Biol.* 13(3):e1005209

79. Esteban O, Markiewicz C, Blair RW, et al. 2018. *FMRIPrep: a robust preprocessing pipeline for functional MRI*. Work. Pap.

80. Eklund A, Nichols TE, Knutsson H. 2016. Cluster failure: Why fMRI inferences for spatial extent have inflated false-positive rates. *Proc. Natl. Acad. Sci. U. S. A.* 113(28):7900–7905

81. Gronenschild EHBM, Habets P, Jacobs HIL, et al. 2012. The effects of FreeSurfer version, workstation type, and Macintosh operating system version on anatomical volume and cortical thickness measurements. *PLoS One*. 7(6):e38234

82. Miller G. 2006. A Scientist's Nightmare: Software Problem Leads to Five Retractions. *Science*. 314(5807):1856–57

83. Kernighan BW, Plauger PJ. 1978. *The elements of programming style*

84. Raymond E. 1999. The cathedral and the bazaar. *Knowledge, Technology & Policy*. 12(3):23–49

85. Bigdely-Shamlo N, Makeig S, Robbins KA. 2016. Preparing Laboratory and Real-World EEG Data for Large-Scale Analysis: A Containerized Approach. *Front. Neuroinform.* 10: